\documentclass[twocolumn,aps, pra, english]{revtex4-1}

\usepackage{amsmath,graphicx,amssymb}
\usepackage{xcolor}
\usepackage{bbold}

\begin{document}

\title{Role of disorder in super- and subradiance of cold atomic clouds}%

\author{Florent Cottier}
\affiliation{Instituto de F\'{i}sica de S\~{a}o Carlos, Universidade de S\~{a}o Paulo - 13560-970 S\~{a}o Carlos, SP, Brazil}
\affiliation{Universit\'{e} C\^{o}te d'Azur, CNRS, INPHYNI, France}
%\affiliation{Institut Non Lin´eaire de Nice, CNRS, Universit´e de Nice Sophia-Antipolis - F-06560 Valbonne, France, EU}

\author{Robin Kaiser}%
\affiliation{Universit\'{e} C\^{o}te d'Azur, CNRS, INPHYNI, France}

\author{Romain Bachelard}
%\affiliation{Instituto de F\'{i}sica de S\~{a}o Carlos, Universidade de S\~{a}o Paulo - 13560-970 S\~{a}o Carlos, SP, Brazil}
\affiliation{Departamento de F\'{\i}sica, Universidade Federal de S\~{a}o Carlos, Rod. Washington Lu\'{\i}s, km 235 - SP-310, 13565-905 S\~{a}o Carlos, SP, Brazil}

%\author{Robin Kaiser}%
%\affiliation{Institut Non Lin´eaire de Nice, CNRS, Universit´e de Nice Sophia-Antipolis - F-06560 Valbonne, France, EU}

\date{\today}

%%%%%%%%%%%%%%%%%%%%%%%%%%%%%%%%%%%%%%%%%%%%%%%%%%%%%%%%%%%%%%%%%%%%%%%%%
% Abstract
%%%%%%%%%%%%%%%%%%%%%%%%%%%%%%%%%%%%%%%%%%%%%%%%%%%%%%%%%%%%%%%%%%%%%%%%%

\begin{abstract}
The presence of superradiance and subradiance in microscopic and mean-field approaches to light scattering in atomic media is investigated. We  show that these phenomena are present in both descriptions, with only minor quantitative differences, so neither rely on disorder. In particular, they are most prominent in media with high resonant optical depth yet far-detuned light, i.e.. in the single--scattering regime.
\end{abstract}

\maketitle
%\tableofcontents

%%%%%%%%%%%%%%%%%%%%%%%%%%%%%%%%%%%%%%%%%%%%%%%%%%%%%%%%%%%%%%%%%%%%%%%%%
% Introduction of the paper
%%%%%%%%%%%%%%%%%%%%%%%%%%%%%%%%%%%%%%%%%%%%%%%%%%%%%%%%%%%%%%%%%%%%%%%%%

\section{Introduction}

Light scattering by disordered samples is usually studied with two kinds of approaches. On the one hand, mean-field (MF) theories are important to understand the coherent propagation of light on the macroscopic scale. One of the most fundamental representation of matter in this context is that of a dielectric, where the atomic details are neglected. In the context of linear optics, this approach allowed for the development of geometrical optics, Mie scattering techniques~\cite{Mie1908, vandeHulst1957}, transfer matrix theory~\cite{Born1999} and many other common tools. Such mean-field treatments can even be refined to higher orders to include, for example, the effect of quantum statistics for the scatterers~\cite{Morice1995,Bons2016}.

On the other hand, microscopic approaches offer an ab-initio treatment which highlights the cooperation of the atoms as they interact through the vacuum light modes~\cite{Lehmberg1970}, as well as the role of disorder and correlations in this context. It is a natural modeling to study few-body scattering, yet for macroscopic samples the associated theoretical and numerical calculations can be daunting even in the linear optics regime, since the macroscopic symmetries are broken by the disorder. A microscopic approach is nevertheless fundamental to understand the role of disorder, and to understand phenomena such as Anderson localization which describes the transition from a conductor to an insulator~\cite{Anderson1958, Abrahams1979}.% or weak localization~\cite{Kuga1984, Albada1985,Wolf1985,Akkermans1986}.

In this context, thanks to the relative absence of decoherence mechanisms, cold atoms represent a powerful platform to investigate the role of the correlations which appear between the scatterers. Cold clouds were consequently used to study phenomena where the interferences between the field of the atomic dipoles play an important role, such as Anderson localization of light~\cite{Skipetrov2014,Bellando2014} on the microscopic side, or superflash~\cite{Kwong2014} and Mie resonance modes~\cite{Bachelard2012} on the mean-field side. The microscopic and mean-field approaches have been shown to provide equivalent results on the presence of polaritonic modes in these clouds~\cite{Jennewein2016, Schilder2016}. Optical forces~\cite{Bienaime2013,Matzliah2017} have also drawn some attention recently, yet the role of dipole-dipole correlations have been discussed to be minor~\cite{Bachelard2016}.  Recently, the microscopic theory emerged as an important tool to study the deviations from mean-field theory due to strong particle-particle correlations, as they appear in the high density regime~\cite{Ruostekoski2016,Javanainen2016} or close to the atomic resonance~\cite{Schilder2017}. 
 
The dynamical phenomena of superradiance (SR) and subradiance have however escaped a detailed microscopic versus mean-field comparison. Dicke superradiance if often considered as the hallmark of the cooperative phenomena, with a coherent coupling between all the atoms~\cite{Dicke1954}: Using a microscopic description, a fully inverted ensemble of two-level atoms was originally shown to present a series of accelerated emission rates. In the linear optics regime, the effect is also known as ``single-photon superradiance'' and is associated to a single accelerated rate~\cite{deOliveira2014,Araujo2016,Roof2016}. The emergence of these short lifetimes scattering modes comes with that of long lifetime modes called subradiant. Light is retained by these atomic modes on times potentially much longer than the single atom excited lifetime. It is different from the incoherent process of radiation trapping, and it has also been reported in the linear optics regime~\cite{Bienaime2012,Guerin2016}. While superradiant modes are generally associated to plane wave phase profile~\cite{Scully2006}, subradiant modes appear better addressed by more complex phase profiles~\cite{Araujo2017}. This rises the important question of the importance of disorder to reach superradiance and subradiance, i.e., are these effects associated to the medium disorder, or are they present in mean-field theories?
 
To address this question, we investigate in this paper these two dynamical phenomena in large atomic clouds, in the linear optics regime, using both microscopic and MF theories. After introducing two models, one based on microscopic scatterers and another based on Maxwell equations in a dielectric medium, we compute the scattering eigenmodes and their dynamics, and show that superradiant and subradiant modes are present in both approaches, with only minor contributions from disorder. A connection between the Mie modes and the superradiant rates is established, whereas subradiant modes are shown to correspond to higher-order scattering modes with a complex phase profile. 

The paper is organized as follows: In Sec.\ref{Sec:models}, the microscopic model based on point--like dipoles and the mean-field one derived from Maxwell equations are presented, and their intrinsic differences discussed.  In section \ref{Sec:spectra}, we compare their scattering eigenmodes and eigenvalues, and compare the superradiant and subradiant lifetimes present in both models. In Section \ref{Sec:emission} the decay dynamics of a cloud is computed, which allows to quantify the observed decay rates. We draw our conclusions on the similarities and differences between the two approaches in Section~\ref{Sec:ccl}.

%%%%%%%%%%%%%%%%%%%%%%%%%%%%%%%%%%%%%%%%%%%%%%%%%%%%%%%%%%%%%%%%%%%%%%%%%
% The two models
%%%%%%%%%%%%%%%%%%%%%%%%%%%%%%%%%%%%%%%%%%%%%%%%%%%%%%%%%%%%%%%%%%%%%%%%%

\section{Mean-field and microscopic models\label{Sec:models}}

\subsection{Derivation from Maxwell equations}

Let us first consider the atomic medium as a dielectric, where the light propagation is ruled by Maxwell equations. For a medium without free charges ($\nabla.\mathbf{E}=0$), the propagation of the electric field $\mathbf{E}$ generated by the electronic current is given by
\begin{equation}\label{WaveEquation}
\nabla^2\mathbf{E} - \frac{1}{c^2}\frac{\partial ^2\mathbf{E}}{\partial t^2}= \mu_0 \frac{\partial \mathbf{j}}{\partial t},
\end{equation}
where $\mathbf{j}(\mathbf{r},t)=-e\rho(\mathbf{r})\mathbf{v}_e$ is the electronic current ($\rho$ being the electronic density), $c$ the speed of light and $\mu_0$ the vacuum permeability. We use a classical description of the current generated by the valence electrons which behave as classical oscillators driven by the electric field. Calling $\mathbf{r}_e(\mathbf{r})$ the displacement of electrons at $\mathbf{r}$ from their equilibrium position, one obtains:
\begin{equation}\label{ElectronDyna}
\frac{\partial^2 \mathbf{r}_e}{\partial t^2} + \omega_a^2 \mathbf{r}_e = -\frac{e}{m} \mathbf{E}(\mathbf{r},t), 
\end{equation}
with $e$ and $m$ the electron charge and mass, and $\omega_a$ the natural frequency of the oscillator. Considering a single direction $\hat{\epsilon}$ for the electronic motion and field, the slowly-varying variables $E$ and $r_e$ can be introduced:
\begin{eqnarray}\label{eq:SlowlyVar}
\mathbf{E}(\mathbf{r},t)&=&E(\mathbf{r},t) e^{-i\omega t}\hat{\epsilon},
\\ \mathbf{r}_e(\mathbf{r},t)&=&r_e(\mathbf{r},t) e^{-i\omega t}\hat{\epsilon},
\end{eqnarray}
which turns Eq.\eqref{WaveEquation} into 
\begin{equation}
\nabla^2 E+ k^2E= \mu_0 e\omega^2 \rho r_e,\label{eq:Ere}
\end{equation}
with $k=\omega/c$. From Eq.\eqref{eq:Ere} the electric field can be obtained from the electronic motion as
\begin{equation}
E(\mathbf{r},t) = -\frac{i\mu_0 e\omega^3}{4\pi c}\int \mbox{d}\mathbf{r}' \rho(\mathbf{r}')G(\mathbf{r}-\mathbf{r}') r_e(\mathbf{r}',t),\label{eq:reE}
\end{equation}
with $G(\mathbf{r}-\mathbf{r}')=\exp(ik|\mathbf{r}-\mathbf{r}'|)/(ik|\mathbf{r}-\mathbf{r}'|)$ the retarded Green's function. Consequently, the dynamics of the electric field is given by:
\begin{equation}
\frac{\partial E(\mathbf{r},t)}{\partial t} =-\frac{i\mu_0 e \omega^3}{4\pi c}\int \mbox{d}\mathbf{r}'\rho(\mathbf{r}') G(\mathbf{r}-\mathbf{r}') \frac{\partial r_e(\mathbf{r}',t)}{\partial t}.\label{eq:reEdt}
\end{equation}
The equation for the field is closed considering the electronic motion \eqref{ElectronDyna} and the slowly-varying approximation \eqref{eq:SlowlyVar}, which leads to
\begin{equation}\label{eq:Eve}
(\omega^2-\omega_a^2)r_e+2i\omega \frac{\partial r_e}{\partial t}=\frac{e}{m}E.
\end{equation}
Introducing $\Delta=\omega-\omega_a$ the detuning between the driving field and the electronic oscillator frequency, and assuming $\Delta\ll\omega_a$, Eqs.(\ref{eq:reE}--\ref{eq:Eve}) allow to derive a self-consistent equation for the electric field~\cite{Svidzinsky2010}:
\begin{equation}\label{eq:EMF}
\frac{\partial E(\mathbf{r},t)}{\partial t} = i\Delta E(\mathbf{r},t)-\Gamma\int \mbox{d}\mathbf{r}'\rho(\mathbf{r}') G(\mathbf{r}-\mathbf{r}') E(\mathbf{r}',t),
\end{equation}
with $\Gamma=\mu_0 e^2 \omega_a^2/8\pi mc$. In this semiclassical picture, $\Gamma$ corresponds to a decay rate for the electric field. In the next section, the microscopic derivation will show that it can be associated to the decay rate of an atomic transition close to resonance.

Let us remark that, in order to avoid a divergence at the origin, the scalar Green function $G$ in Eq.\eqref{eq:EMF} must be defined as
\begin{equation}
G(\mathbf{r}) = \left\{ 
\begin{array}{cl} 
 \exp(\imath kr)/(\imath kr) & \mbox{ for $r>0$},
 \\ 1 &\mbox{ for $r=0$} .
\end{array} 
\right.
\end{equation}
On the one hand, this definition at the origin allows to recover the single-atom decay when ones moves to a microscopic description, using a density of the form $\rho(\mathbf{r})=\sum_{j=1}^N\delta(\mathbf{r}-\mathbf{r}_j)$. On the other hand, while the real part of $G(r)=-\imath\cosh(kr) + \sinh(kr)$ converges to $1$ as $r\to0$, its imaginary part diverges. This singularity is at the origin of the Lamb shift, and it can be addressed using renormalization techniques~\cite{Scully1997}. In the present work, it is put to zero, assuming the shift is already included in the transition frequency. All in all, the three-dimensional integral of $G$ around the origin, for a smooth density $\rho$ and field $E$, gives a finite contribution, so the divergence is purely local and does not require the introduction of any cut-off when performing, e.g., numerical integrations.

This continuous approach corresponds to a mean-field treatment of the cloud, which is in particular at the basis of the dielectric description of matter. Indeed in the stationary regime the dipole field can be shown to obey Helmholtz equation, so techniques such as Mie scattering can be used to describe the scattering properties of the atomic sample~\cite{Bachelard2012}. In this context, the study of higher-order terms (in the stationary state) leads to corrections such as the Lorentz--Lorenz equation~\cite{Born1999} but also effects related to quantum statistics for degenerate gases~\cite{Morice1995,Bons2016}.

\subsection{Microscopic derivation}

From a microscopic point of view and using a dipole approximation, our medium is composed of an ensemble of $N\gg1$ point scatterers, which are here modelled as two-level atoms (polarizations effects are neglected). Their position $\mathbf{r}_j$ is here considered to be fixed, and $\Gamma$ refers to the linewidth and $\omega_a$ to the frequency of the atomic transition. The system is initially pumped with a monochromatic plane wave of wavevector $\mathbf{k} = k \mathbf{\hat{z}}$, detuned from the atomic transition by $\Delta = \omega - \omega_a$, and with Rabi frequency $\Omega$. In the low-intensity regime ($\Omega \ll \Gamma$) and using the Markov approximation, the resonant dynamics of the atomic dipoles is given by a set of $N$ coupled equations~\citep{Lehmberg1970,Courteille2010} for the atomic dipoles $\beta_j$:
\begin{eqnarray}
\frac{d\beta_j}{dt} &=& \left( \imath\Delta-\frac{\Gamma}{2} \right) \beta_j - \frac{\imath \Omega}{2} e^{\imath\mathbf{k .r}_j} \nonumber
\\ &&- \frac{\Gamma}{2} \sum_{m \neq j} \frac{\exp(\imath k|\mathbf{r}_j-\mathbf{r}_m|)}{\imath k|\mathbf{r}_j-\mathbf{r}_m|}\beta_m.\label{ODEMicro}
\end{eqnarray}
%The first right-hand term describes the single atom dynamics, with both spontaneous emission and oscillations due to the detuning, the second term corresponds to the pump, and the last one to the radiation from all other atoms.
This problem of linear optics is thus characterized by the matrix $D$ which describes the coupling between the dipoles: the diagonal terms $D_{i,i} = (1-2\imath \Delta/\Gamma)$ correspond to the single-atom dynamics, the off-diagonal ones $D_{i,j\neq i} = \exp(\imath k|\mathbf{r}_i-\mathbf{r}_j|)/(\imath k|\mathbf{r}_i-\mathbf{r}_j|)$, to the exchange of photons between the dipoles. This coupling between the atoms leads to the emergence of collective scattering modes, given by the eigenvectors $\hat{\psi}_n$ of $D$, whose components $\hat{\psi}_n^j$ correspond to the contribution of atom $j$. The real part of the associated eigenvalue $\lambda_n$ corresponds to the decay rate (in units of $\Gamma$) of the mode, and the imaginary part to its energy (relative to the atomic transition).

In the linear optics regime the dipole $\beta_j$ is directly proportional to the local electric field $E_j$ ($\beta_j=dE_j/\hbar(2\Delta+i\Gamma)$, with $d$ the matrix dipole element of the transition, such that $\Omega=dE/\hbar$). Thus Eq.\eqref{ODEMicro} can be seen as a microscopic version of Eq.\eqref{eq:EMF}, up to two terms. First, the single-atom decay term is not present in the MF model, since during its derivation the medium was considered to be composed of a smooth density (field approach), rather than point scatterers. Second, the pump term is absent in the MF model: While it is irrelevant for the analysis of the eigenmodes and eigenvalues of the system, it will be added by hand in the section where the decay dynamics is studied, to understand how a plane wave couples to the system.

The study of the scattering eigenmodes of $D$ has been the object, in the steady-state regime, of several statistical studies~\cite{Skipetrov2011,Goetschy2011,Goetschy2011b}, in particular in the context of Anderson localization of light~\cite{Skipetrov2014,Bellando2014} which emerges due to the disorder in the atomic dipole distribution. In the microscopic model, the eigenvalues of the stationary and of the decay problem are the same, so a direct connection can be drawn before static problems such as Anderson localization, and dynamical ones such as subradiance. On the contrary, the MF model has different sets of eigenvalues in each case, so the decay problem possesses its own equation: Differently from the stationary case, the dynamical equation presents an implicit equation for the eigenvalues~\cite{Svidzinsky2010}. 

In order to understand the potential consequences of this difference in the structure of the equations, and on the role of disorder more generally, we now turn to characterizing the subradiant and superradiant eigenvalues in each model.

\section{Scattering modes\label{Sec:spectra}}

Since the scattering problem is a linear one (that is, in the linear optics regime), many of its properties can be deduced by studying the properties of the matrix which describes the coupling between the atoms. This section is dedicated to comparing the scattering eigenmodes of the two models, by direct diagonalization.

The microscopic problem generates a coupling matrix of size $N\times N$, which allows for diagonalization procedures to address exactly systems of up to a dozen thousands particles.

Differently, the MF model, however, is infinite-dimensional, and thus requires some simplifying hypotheses to become tractable. One of them is to assume a spherical symmetric for the density $\rho(r)$~\cite{Mie1908,vandeHulst1957},  in which case the spherical harmonics $Y_{n,m}$ appear as a natural basis to tackle with it. Indeed the retarded Green function is diagonal in this basis:
\begin{eqnarray}\label{coupling}
 \frac{e^{ik|\textbf{r}-\textbf{r}'|}}{\imath k|\textbf{r}-\textbf{r}'|} &=& 4\pi \sum_{n=0}^\infty \sum_{-n\leq m\leq n} Y_{n,m}(\theta,\phi)Y_{n,m}^*(\theta',\phi')\nonumber \\
 & & \left\{ 
\begin{array}{ll} 
 j_n(k r')h_n(k r) & \mbox{ for $r > r'$},\\
 j_n(k r)h_n(k r') &\mbox{ for $r \leq r'$},
\end{array} 
\right.
\end{eqnarray}
where $j_n$ and $h_n$ are respectively the spherical Bessel and Hankel functions, and where the polar angle is chosen with reference to the wavevector $\mathbf{k}=k \mathbf{\hat{z}}$. In that basis the dipole field and the incident plane wave decompose as
\begin{eqnarray}\label{eq:decbeta}
\beta(\mathbf{r},t) &=& \sum_{n,m} \hat{\beta}_n(r,t)Y_{n,0}(\mathbf{\hat{r}}),
\\ \label{eq:decpw} e^{i\mathbf{k}_0.\mathbf{r}} &=& \sqrt{4\pi} \sum_n \sqrt{2n+1}i^n j_n(k r)Y_{n,0}(\theta,\phi).
\end{eqnarray}
Injecting (\ref{eq:decbeta}--\ref{eq:decpw}) into \eqref{eq:EMF} and projecting the result on the $(n, m=0)$ spherical harmonics provides the dynamical equations for $\beta_n$:
\begin{eqnarray}\label{ODEFluidn}
 &&\frac{d\beta_n(r,t)}{dt} =  i\Delta \beta_n(r,t) - E_0 \sqrt{4\pi}\sqrt{2n+1}i^n j_n(k r) \\ \nonumber
 & & - 2\pi\Gamma \int_0^R \,dr' \rho(r')\beta_n(r',t)  \left\{ 
\begin{array}{ll} 
 j_n(k r')h_n(k r) r'^2 & \mbox{for $r > r'$}\\
 j_n(k r)h_n(k r') r'^2 &\mbox{for $r \leq r'$} 
\end{array} 
\right.
\end{eqnarray} 
Since polarization effects are neglected, the $m\neq0$ modes are not populated by a plane wave as the rotational symmetry around $\mathbf{\hat{z}}$ is preserved for the MF model. Anyway the modes $(n,m\neq0)$ obey the same equation as the $(n,0)$ one, so each eigenvalue of \eqref{ODEFluidn} has a degeneracy $(2n+1)$. In the microscopic case, the symmetry is broken by the atomic disorder.

The eigenvalues $\lambda_n$ of the MF problem, which are associated to modes of the form $\beta_n(r,t)=\exp(-\lambda_n \Gamma t)f_n(r)$, are given formally by an implicit equation~\cite{Svidzinsky2010}. As the problem is infinite-dimensional, the set of eigenvalues is infinite. Consequently finding the eigenvalues by some root algorithm (based on Newton's method, for example) is inefficient, since the eigenvalues found may not be representative of the overall set. Hereafter, in order to compute the spectrum and the dynamics of the radial equation~\eqref{eq:EMF}, we project it on a grid of step $h$ and with a finite domain of definition. In this work we consider Gaussian atomic densities of rms radius $\sigma_r$ (i.e., $\rho(r)=N\exp(-r^2/2\sigma_r^2)/(\sqrt{2\pi}\sigma_r)^3$), and an integration range $[0; R=3\sigma_r]$.

The projection of the radial problem on a grid brings us back to a finite matrix, of size $H\times H \sim R/h \times R/h$, which is diagonalized numerically to obtain a set of eigenvalues  $\lambda_{n,j}$, $j=1,..H$ for each mode $n$. The accuracy of this finite set of eigenvalues, as compared to the continuous one, is evaluated by decreasing the step $h$ and observing its convergence. Practically, decreasing $h$, and thus inversely proportionally increasing $H$, generates more eigenvalues, which are smaller and smaller (see discussion below); the largest eigenvalues of the spectrum appear to be accurately computed for $h=3\lambda/100$, which is value used throughout this work. %We remind that the imaginary part of the diagonal term is put to $0$ to avoid the divergence due to the interaction of the atom with himself.

This fundamental difference between the microscopic and continuous approach, i.e. finite versus infinite set of eigenvalues, has an important consequence. Indeed, the trace of the coupling matrix $D$ is the same for both systems (i.e., $N(1-2i\Delta/\Gamma)$. Thus, in the microscopic model which has $N$ eigenvalues, the coupling terms $D_{i,j\neq i}$ leads to the simultaneous emergence of both superradiant ($\Re(\lambda_n)>1$) and subradiant ($\Re(\lambda_n)<1$) scattering modes. 
%We now look at the same property for the continuous model whereas it has an infinite number of eigenvalues. Using Eq.\ref{ODEFluid n}, we have :
%\begin{equation}\label{EqSumSum}
%\begin{array}{lll}
%\sum_{n,j} \lambda_{n,j} &=& -\sum_{n=0}^{\infty} \sum_{j=1}^{H} \left[\imath \Delta - \frac{\Gamma}{2} 4\pi h^3 \rho(jh) \right.\\
%  & \times & \left. \Re\left( j_n(kjh)h_n(kjh) \right) \times (2n+1) \right]
%\end{array}
%\end{equation}

\begin{figure}[!h]
%\centering
%\hspace{-1cm}
\includegraphics[scale=.7]{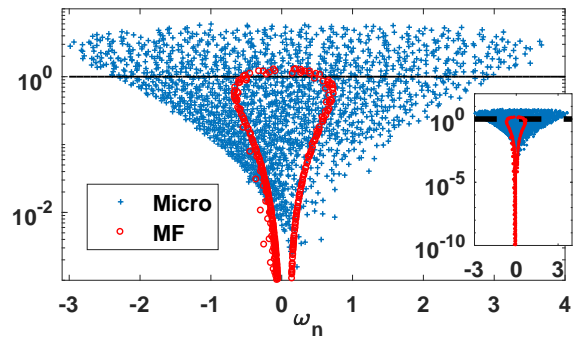} 
\caption{\label{EigenvaluesComplexSpace}Eigenvalues in the complex plane $\lambda_n=\gamma_n+i\omega_n$ for the microscopic (blues crosses) and continuous (red circles) models. Simulations realized with a Gaussian cloud of $N = 2000$ particles and $\sigma_r\approx12$ ($b_0\approx 28.4$), at resonance ($\delta = 0$). }
\end{figure}

On the contrary, in the MF model the presence of an infinity of eigenvalues with positive real part, but a finite trace, only guarantees the presence of an infinite series of arbitrarily low decay rates $\Re(\lambda_n)$, while superradiance is not guaranteed. This point is illustrated in Fig.\ref{EigenvaluesComplexSpace}, where an example of sets of eigenvalues are presented. Both superradiant and subradiant modes appear in the two models, but the superradiant rates are lower in the MF case. In the inset one can also observe the trail of eigenvalues of the MF model, which is limited only by numerical precision and by the grid step $h$. 

Let us comment that throughout this work the simulations are realized below the Anderson localization threshold~\cite{Skipetrov2014,Bellando2014} ($\rho\lambda^3 \approx 22$), so the subradiant modes observed are not associated to exponentially localized modes but rather to subradiance in dilute atomic samples~\cite{Bienaime2012,Guerin2016}.

\begin{figure}[!h]
\centering
\includegraphics[scale=0.8]{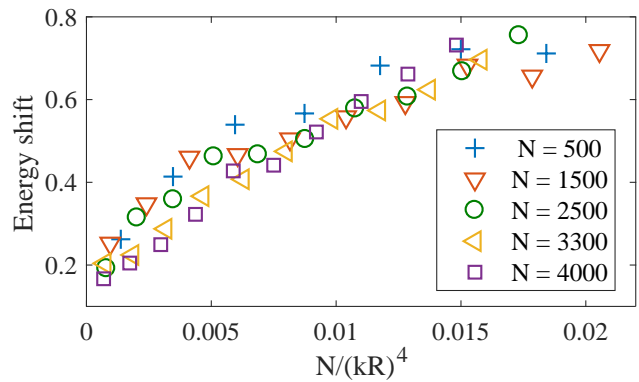} 
\caption{\label{EnergyShiftShift}Energy shift of the subradiant tail, for a cloud with uniform density, and illuminated at resonance. The energy shift is computed using the average energy shift of the $5\%$ most subradiant eigenvalues.}
\end{figure}

In the microscopic case, the most subradiant eigenvalues presents a shift in energy, which is absent from the mean-field model. We evaluate it by computing the average energy of the $5\%$  most subradiant ones: It corresponds to the laser frequency which optimizes the population of the most subradiant modes. As shown in Fig.\ref{EnergyShiftShift} the shift scales as $N/(kR)^4$, differently from the Lorentz-Lorenz and collective Lamb shifts which scale with the density~\cite{Friedberg2010}. This specific scaling suggests that for large samples and in the low density limit, the shift will vanish, contrary to cooperative phenomena which scale as the resonant optical thickness $b_0=2N/(k\sigma_r)^2$~\cite{Guerin2016b}.

Scattering modes with long lifetimes are often associated with spatial localized, as in the case of Anderson localization or in whispering gallery modes. These latter modes are not present in our system due to the Gaussian density considered: Smooth densities do not allow for surface modes to propagate by internal reflection~\cite{Bachelard2012}. 
As for long lifetimes modes inside the cloud, they have been shown to be responsible, in the microscopic model of scalar light, for a localization transition~\cite{Skipetrov2014,Skipetrov2016}. 

\begin{figure}[!h]
%\centering
\includegraphics[scale=0.65]{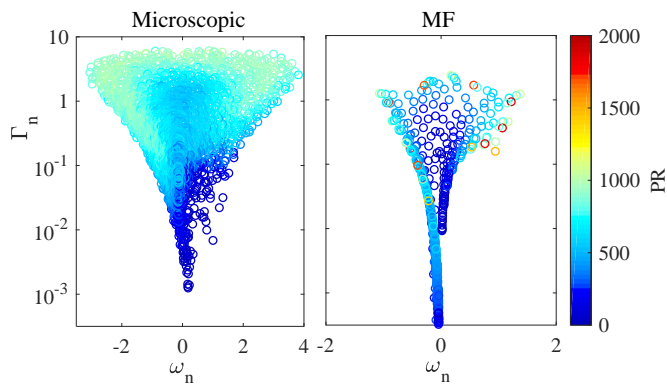} 
%\\ \includegraphics[scale=0.09]{Sigma_R_Micro.jpg} 
\caption{
\label{PR}Participation ratio of the microscopic (left) and MF (right) models for a Gaussian cloud with $N = 2000$ atoms with optical thickness $b_0\approx 30$, at resonance.
}
\end{figure}

Let us first check the connection between the lifetime of the modes and their spatial extension of the modes by computing their participation ration (PR), which quantifies the number of atoms participating substantially to it. The PR of mode $n$ is defined in, resp., the microscopic and MF model as:
\begin{eqnarray}\label{IPRMicro}
\text{PR}_n &=& \frac{\left(\sum_j |\beta^n_j|^2\right)^2}{\sum_j |\beta^n_j|^4},
\\ \text{PR}_n &=& \frac{\left( \int \rho(\mathbf{r})|\beta_n (\mathbf{r})|^2d\mathbf{r}\right)^2}{\int\rho(\mathbf{r}) |\beta_n (\mathbf{r})|^4 \,d\mathbf{r}}
\end{eqnarray}
%After projection on the spatial grid we obtain for each eigenvalue $\lambda_{n,j}$ :
%
%\begin{equation}\label{IPRFluid}
%IPR^{n,j} = \frac{\sum_i\rho(r_i)|\beta^i_{n,j}|^4 r_i^2 \,\, \int_{\Omega} |Y_{n,0}(\hat{r})|^4\,d\Omega}
%{\left(\sum_i\rho(r_i)|\beta^i_{n,j}|^2 r_i^2\right)^2}
%\end{equation}
Fig.\ref{PR} depicts the PR for both models and the PRs are qualitatively similar: Superradiant mode have a large PR, which correspond to spatially extended modes; longer lifetime are associated to a lower PR, and thus to a stronger localization in space. Note that, as the mean-field approach contains no disorder, one does not expect a disorder-based localization transition, with the emergence of exponentially localized modes.

%%%%%%%%%%%%%%%%%%%%%%%%%%%%%%%%%%%%%%%%%%%%%%%%%%%%%%%%%%%%%%%%%%%%%%%%%
% Super and subradiance scaling 
%%%%%%%%%%%%%%%%%%%%%%%%%%%%%%%%%%%%%%%%%%%%%%%%%%%%%%%%%%%%%%%%%%%%%%%%%
\section{Radiated intensity\label{Sec:emission}}

Although the eigenmodes appear to have similar features in the two models, it reveals only part of the information on the cloud radiation. Indeed it does not tell how a plane--wave couples to each mode~\cite{Guerin2017}. In particular, a remarkable aspect of the coupled dipole problem is that in the large detuning, despite the cloud becomes essentially transparent to the incident light, superradiance and subradiance are still present. Furthermore, these two phenomena present a scaling with the resonant optical thickness $b_0=2N/(k\sigma_r^2)$~\cite{Bienaime2012,deOliveira2014,Guerin2016,Araujo2016,Roof2016,Guerin2016b}. It is  different, for example, from radiation trapping which scales with the optical thickness $b=b_0/(1+4\delta^2)$ and vanishes in the large detuning limit ($b\to 0$).

\begin{figure}[!h]
    \centering
        \includegraphics[scale=0.45]{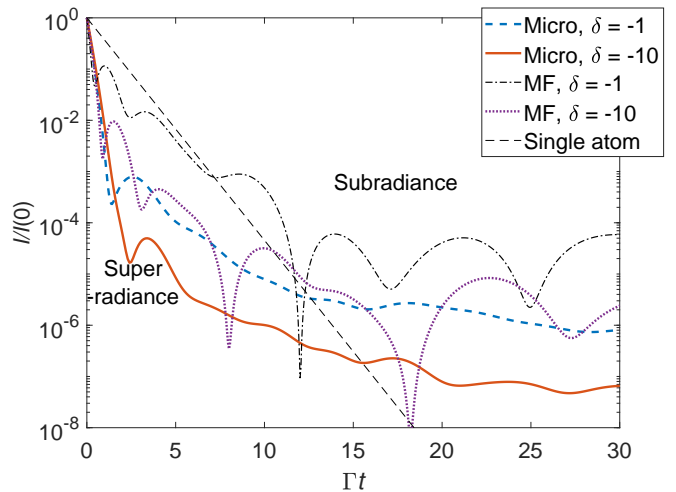} 
\caption{\label{Isub}Emission dynamics of the forward scattered light after the laser is switched off ($t=0$), for the microscopic and the MF models. Simulations realized for a Gaussian cloud with $b_0\approx 28$ and $N = 1900$ atoms, close ($\delta=-1$) and far ($\delta=-10$) from resonance.}
\end{figure}

To understand which modes couple to an incident plane-wave, we here compare the dynamics of the radiated intensity in both models when the incident laser, after a long period on during which the system was charged, is abruptly cut. 
The intensity radiated by the ensemble of dipoles in direction $\mathbf{n}$ and in the far field limit writes, for the microscopic and the MF models respectively: 
\begin{eqnarray}\label{IntensityMicro}
I(t) &\propto& \left| \sum_{j=1}^N \beta_j(t) e^{-\imath k\mathbf{n}.\mathbf{r_j}} \right|^2,
\\ I(t) &\propto& \left| \int_V \rho(\mathbf{r}) \beta(\mathbf{r},t) e^{-\imath k\mathbf{n}.\mathbf{r}} \,d^3\mathbf{r} \right|^2.
\end{eqnarray}
As can be observed in Fig.\ref{Isub} the emission of the cloud is first characterized by the fast decay associated to the superradiant modes, which dominates the short time dynamics as they initially carry most of the energy. Yet they quickly lose their population and soon the long-lived emission of the subradiant modes takes over. Interestingly, both models present super- and subradiance, including for large detuning where the single-scattering regime is reached ($\delta=-10$ curves, corresponding to an optical thickness of $b\sim0.07$). The two models also present significant oscillations of the intensity during the decay dynamics, when many modes with similar energies compete.

\begin{figure}[!h]
        \includegraphics[scale=0.4]{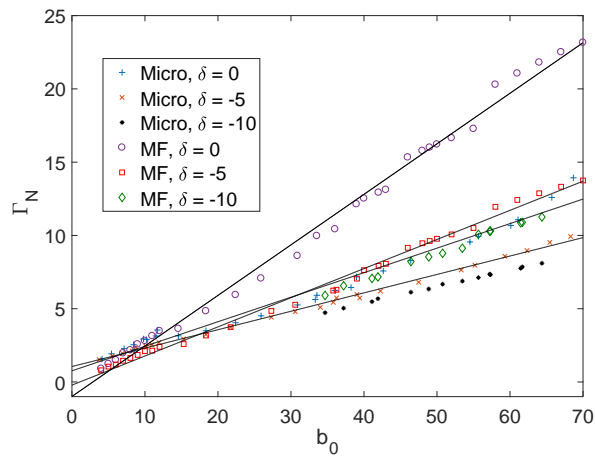}
\caption{\label{GammaN_delta}SR rate in the forward direction ($\theta=0$) as a function of the resonant optical thickness for the microscopic and mean-field models, at resonance ($\delta=0$) and out of resonance ($\delta=-5,\ -10$). Simulations realized for a spherical Gaussian cloud of $N=1000$ atoms.}
\end{figure}

A detailed study of the dependence of the superradiant rate on the cloud characteristics (size and particle number) and on the light-atom coupling (detuning) reveals that, just as the microscopic approach, the relevant scaling parameter is the {\it resonant} optical thickness (see Fig.\ref{GammaN_delta}). More specifically, the SR rate scales linearly with $b_0$. Consequently, the phenomenon is still present in the large detuning limit, only their population is decreased (and thus the radiated intensity) as the cloud turns transparent.

This confirms that the continuous model captures accurately the cooperativity of superradiance, despite a discrepancy of a factor $\sim2$ between the two models. Interestingly, the microscopic model shows an {\it enhanced} SR rate compared to the fluid mode, which suggests disorder is actually favorable to SR.

Let us finally comment that the continuous model presents a SR rate of $\Gamma_{SR}\to 0$ in the low-$b_0$ limit: Contrary to the microscopic model where single-atom physics is recovered in the very dilute limit ($\Gamma_{SR}\approx\Gamma$), the continuous limit is not appropriate to describe this limit. 
%Indeed it fails to recover the limit of a single point scatterer with vanishing interactions with his neighbors.

%This result suggests that the superradiance is well captured by the continuous model, which describes the coherent light scattering. Superradiance was originally introduced as a coherent process in the sense that a fully inverted system was shown to decay through a series of symmetric states, with the associated accelerated superradiant rates. Although, it may be tempting to conclude that superradiance is associated to coherent light scattering. \\
%We also oberve, Fig.\ref{GammaN_delta}, that at resonance ($\delta \approx 0$), we have a different behaviour what was already observed \cite{MichelleSuper},  \citep{MichelleSub}. Indeed, at resonance there is a strong absorption and the diffused light predominates which is not described by the continuous model and explains it. \\

%%%\begin{figure}[!h]
%%%\centering
%%%\includegraphics[scale=0.2]{AlldeltaFluid_N10000_step100} 
%%%\caption{\label{TauxSuperFluidDetuning}Superradian rate in the forward direction with a Gaussian cloud of atoms and a scalar light, $b_0 = 2N/(kR)^2$ and $N = 10000$ particles. I used $100 R/\lambda$ points for the spacial grid.}
%%%\end{figure}

Superradiance corresponds to a coherent emission by in-phase dipoles. This typically leads to looking for SR in the direction of the coherent driving, where a coherent phase pattern is imprinted~\cite{Scully2006}. Yet, it has been shown recently that the off-axis emission was also superradiant~\cite{Araujo2016} despite the atomic dipoles are not expected to exhibit specific phase pattern in the transverse directions, where diffuse light is emitted.

\begin{figure}[!h]
\centering
\includegraphics[scale=0.8]{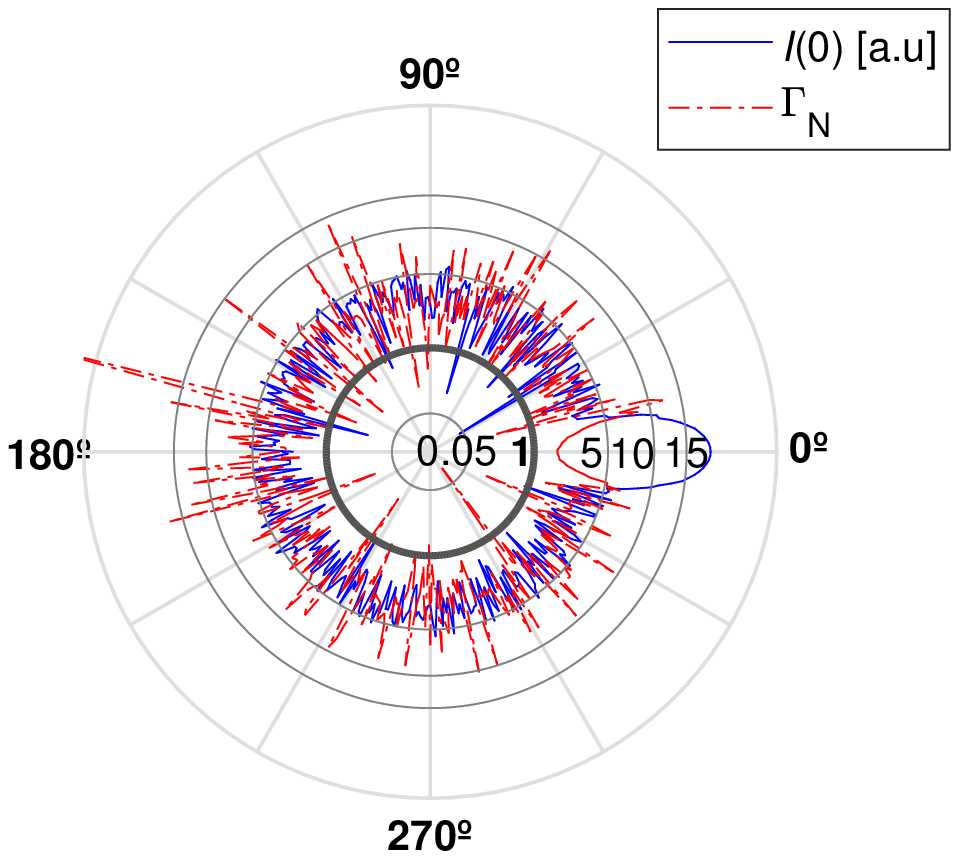}
\includegraphics[scale=0.8]{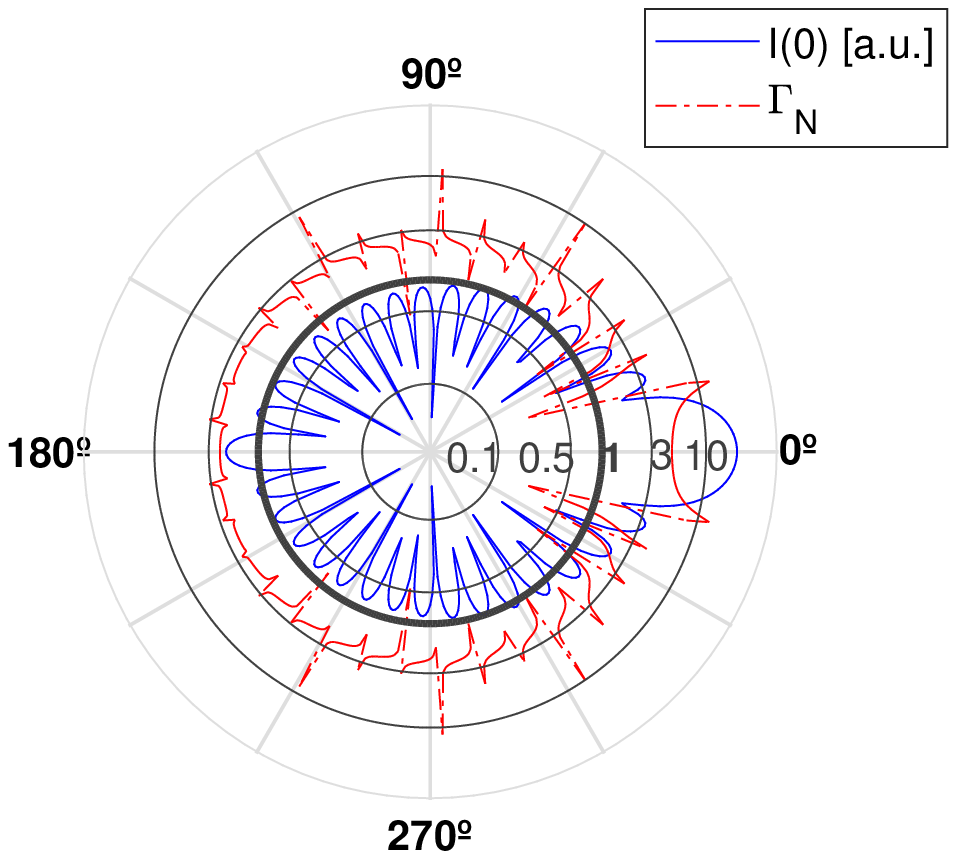} 
\caption{\label{I0_Gamma_theta} Angular dependence of the initial radiated intensity $I(0)$ (blue plain line) and SR rate $\Gamma_N$ (red dash--dotted line) in logscale for the (a) mean-field and (b) microscopic models. The rate is computed over the time window $t\in[0; 0.1]/\Gamma$ for a cloud charged by a plane-wave during a time $50/\Gamma$ until $t=0$. Simulations realized for a Gaussian cloud with $b_0 = 28.7$, $\delta = -10$ and $N = 1908$. The gray circles describe the level of the SR rate, the thick one corresponding to the single--atom rate $\Gamma_N=1$.}
\end{figure}

Fig.\ref{I0_Gamma_theta}(a) illustrates this point, showing the angular dependence of the short-time emission rate in the microscopic model: Despite strong fluctuations due to the atomic disorder, it is superradiant in almost all directions. At initial times, subradiant rates appear only in arbitrary directions, depending on the specific realization. In particular, as reported previously~\cite{Araujo2016}, SR is slightly weaker in the direction of the driving where more energy is radiated.

The MF model presents similar features, despite it does not have density fluctuations. Instead, the SR rate changes with the direction of observation, and it is in general lower in the direction of the lobes of Mie scattering where more of the energy is radiated. This preservation of the off-axis SR rate in both models is particularly surprising considering that in these directions, for dense clouds the diffuse light dominates in the microscopic model whereas only Mie lobes of coherent scattering are present in the MF model.

\begin{figure}[!h]
\centering
\includegraphics[scale=0.6]{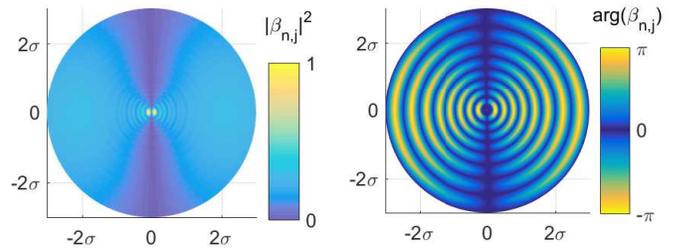}
\caption{\label{fig:SpatialProfile}Spatial (a) intensity and (b) phase profiles of a subradiant mode of the MF model: $\gamma_{n,j}\approx 0.04\Gamma$, corresponding to $(n,j)=(1, 13)$ for a Gaussian cloud with the same parameters as in Fig.\ref{EigenvaluesComplexSpace}.}
\end{figure}

Regarding subradiance, the long--lived modes are also present in the MF model, which can be somehow counter--intuitive: SR modes are associated to in--phase dipoles whereas subradiant modes are associated to less regular phase patterns. Thus, the MF and its macroscopic modes appears as less suitable to support subradiant modes, as compared to the microscopic disordered model. However, higher-order modes in the MF approach (i.e., corresponding to large indexes $n$) correspond to higher-order spherical harmonics, which precisely present spatial pattern with a large number of oscillations (see Fig.\ref{fig:SpatialProfile}). This analysis is also consistent with the fact that the large-$n$ limit is associated to deeply subradiant modes (see inset of Fig.\ref{EigenvaluesComplexSpace}).

A more quantitative description of subradiance requires the characterization of the emission rate at late times ($t>1/\Gamma$). Because of the oscillations of the radiated intensity originating in the interference between several modes with slightly different energy, an averaging procedure is necessary to obtain monotonic decay curves and extract a subradiant rate. Whereas this is performed by averaging over disorder configurations in the microscopic case~\cite{Bienaime2012,Guerin2016}, the intensity in the MF model cannot be averaged easily, as it contains no disorder. Furthermore, creating different ``configurations'' by changing slightly the cloud characteristics (size, density, particle number) does not allow to smooth efficiently the intensity dynamics as the modes (and their energy) is only slightly affected.

\begin{figure}[!h]
\centering
\includegraphics[scale=.7]{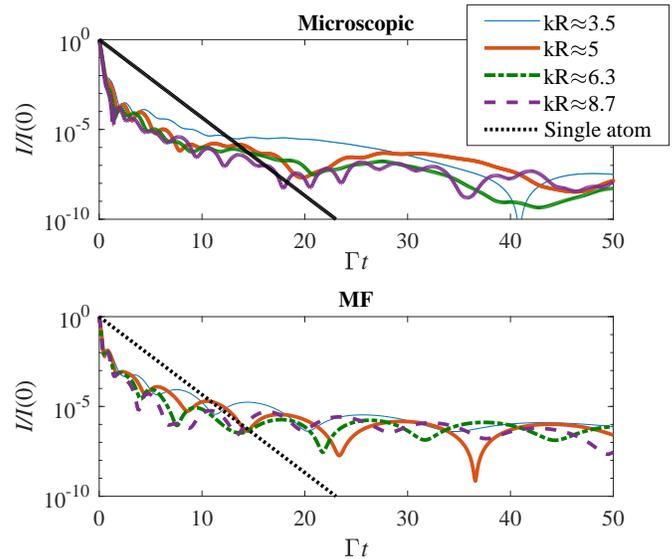} 
\caption{\label{I_k0R} Far-field intensity radiated in the forward direction, for a  Gaussian cloud distribution with $b_0 = 80$ and driven far from resonance ($\delta = -10$).}
\end{figure}

Despite a quantitative comparison is difficult to obtain, the cooperative nature of the subradiance observed in the MF model, as opposed to the incoherent phenomenon of radiation trapping, can be assessed by studying the decay dynamics in the large-detuning limit, where the cloud optical thickness becomes vanishing. Fig.\ref{I_k0R} describes the decay dynamics for the two models, for a single realization and for different values of detunings: Apart from the qualitative resemblance between the two dynamics, an important point is that in both cases the far-detuned limit does not correspond to single-atom dynamics, i.e., collective modes are still present, associated to decay rates very different from the single atom one. This confirms the existence of subradiance in the MF model, in a regime where radiation trapping is marginal.

\section{Discussion \& Conclusion\label{Sec:ccl}}

Our results show that subradiance in dilute cloud does not require disorder, but also exist in an homogeneous medium when higher-order scattering modes are taken into account. These modes correspond to spatial patterns with a strongly oscillating phase. 
Oppositely, superradiant modes, whose decay rate are qualitatively similar in the MF and microscopic approach, correspond to more homogeneous phase profiles (associated to lower-order spherical harmonics). This was expected from typical approaches for superradiance, such as the decay through symmetric states in subwavelength atomic samples~\cite{Dicke1954}, or mean-field approaches for extended samples~\cite{Scully2006,Courteille2010}.

While the recent literature has welcomed several contributions about the deviations from MF approaches observed in dense atomic media~\cite{Javanainen2014,Javanainen2016,Schilder2016, Javanainen2017, Schilder2017} (sometimes called ``standard optics''), our results show that (single-photon) superradiance and subradiance in dilute atomic clouds may instead be compatible with linear optics of polarizable media.

\acknowledgments
R.B. hold Grants from São Paulo Research Foundation
(FAPESP) (Grant Nos. 2014/01491-0 and 2015/50422-4). R.B. and R.K. received support from project CAPES-COFECUB (Ph879-17/CAPES 88887.130197/2017-01). We thank fruitful discussions with W. Gu\'erin.

%%%%%%%%%%%%%%%%%%%%%%%%%%%%%%%%%%%%%%%%%%%%%%%%%%%%%%%%%%%%%%%%%%%%%%%%%
% Bibliography
%%%%%%%%%%%%%%%%%%%%%%%%%%%%%%%%%%%%%%%%%%%%%%%%%%%%%%%%%%%%%%%%%%%%%%%%%

%\newpage

%\bibliographystyle{stylename}
\bibliography{../../Biblio/BiblioCollectiveScattering}

\end{document}